\documentclass[12pt,a4paper]{article}
\usepackage{epsfig,epsf,mathrsfs,amssymb,amsmath}
\usepackage[hang,small,bf]{caption}
\usepackage{amsfonts}
\def\Eqn#1{Eq.~(\ref{#1})}

\def\Eqns#1#2{Eqs.~(\ref{#1}) and~(\ref{#2})}

\def\Fig#1{Fig.~{\ref{#1}}}

\def\Figs#1#2{Figs.~{\ref{#1}} and {\ref{#2}}}

\def\ptmin{p_{T{\rm min}}}

\def\spa#1.#2{\left\langle#1\,#2\right\rangle}
\def\spb#1.#2{\left[#1\,#2\right]}
\def\spaa#1.#2.#3{\langle\mskip-1mu{#1}
                  | #2 | {#3}\mskip-1mu\rangle}
\def\spbb#1.#2.#3{[\mskip-1mu{#1}
                  | #2 | {#3}\mskip-1mu]}
\def\spab#1.#2.#3{\langle\mskip-1mu{#1}
                  | #2 | {#3}\mskip-1mu\rangle}
\def\spba#1.#2.#3{\langle\mskip-1mu{#1}^+
                  | #2 | {#3}^+\mskip-1mu\rangle}

\def\feynsl#1{
  \setbox0=\hbox{/} \setbox1=\hbox{$#1$}
  \dimen0=\wd0 \advance\dimen0 by -\wd1 \divide\dimen0 by 2
  \ifdim\wd0>\wd1 \raise.15ex\copy0\kern-\wd0\kern\dimen0\copy1\kern\dimen0
  \else \kern-\dimen0\raise.15ex\copy0\kern-\dimen0\kern-\wd1\copy1\fi}

\newskip\humongous \humongous=0pt plus 1000pt minus 100pt

\newif\ifdtup

\makeatletter
\def\@eqnnum{\hbox{\reset@font\rm(\theequation)}}
\let\make@eqnnum=\@eqnnum %
\def\eqnum#1{\dec@eqnnum \global\def\make@eqnnum{\reset@font\rm(#1)}%
\def\@currentlabel{#1}%
}
\def\inc@eqnnum{\addtocounter{equation}{1}}
\def\dec@eqnnum{\addtocounter{equation}{-1}}
\@definecounter{equation}%
\@addtoreset{equation}{section} %
\def\theequation@prefix{{\thesection}.} %
\def\theequation{\theequation@prefix\arabic{equation}}%
\makeatother

\begin{document}
\topskip 1cm

\begin{titlepage}
\hspace*{\fill}\parbox[t]{4cm}{
IPPP/02/43\\
DCPT/02/86\\ 27th September, 2002}
\vfill
\begin{center}
{\Large\bf A $Z$-Monitor to Calibrate Higgs Production via Vector Boson
Fusion with Rapidity\\[7pt] Gaps at the LHC}\\
\vspace{1.cm}

{V. A. Khoze$^1$, M. G. Ryskin$^2$, W. J. Stirling$^1$ and P. H.
Williams$^1$}\\
\vspace{.2cm}
{$^1$\sl Institute for Particle Physics Phenomenology\\
University of Durham\\
Durham, DH1 3LE, U.K.}\\

\vspace{.2cm}
{$^2$\sl Petersburg Nuclear Physics Institute\\
Gatchina, St. Petersburg, 188300, Russia}\\

\vspace{.5cm}

\begin{abstract}
  We study central $Z$-boson production accompanied by rapidity gaps on
  either side as a way to gauge Higgs weak boson fusion production at the
  LHC. We analyse the possible backgrounds for the $b\bar{b}$-decay mode and
  show that these can be substantially reduced. Special attention is paid to
  the evaluation of the gap survival factor, which is the major source of
  theoretical uncertainty in the rate of $H$, $Z$ and $W$ central production
  events with rapidity gaps.
\end{abstract}

\end{center}
 \vfill

\end{titlepage}

\section{Introduction}

Hunting the Higgs boson(s) is the highest priority of the international
high-energy physics programme. The Standard Model-like Higgs boson should
have a mass between the LEP2 limit of 114~GeV and the upper bound of about
200~GeV, which is favoured by electroweak data~\cite{1}. Within the MSSM, the
light scalar Higgs boson is expected to be lighter than about 135 GeV, see
for example~\cite{2}. The focus now is on searching for the Higgs at present
and forthcoming hadron colliders, namely the Tevatron and the LHC.\par
To ascertain whether a Higgs signal can be seen, it is crucial to show that
the background does not overwhelm the signal. For instance, the major
difficulty in observing inclusive production of the Higgs in the preferred
mass range around 115~GeV via the dominant $H\rightarrow b\bar{b}$ mode is
the huge $b\bar{b}$ QCD background. In order to rescue the $b\bar{b}$ Higgs
signal different options have been proposed in the literature. An attractive
possibility to reduce the background is to study the central production of
the Higgs in events with a large rapidity gap on either side, see for example
[3--14]. An obvious advantage of the rapidity gap approach is the clean
experimental signature -- hadron free zones between the remnants of the
incoming protons and the Higgs decay products.  The cleanest situation is in
the double-diffractive exclusive process:
\begin{equation}
\label{excl} 
p\overset{(-)}{p}\rightarrow p + H + \overset{(-)}{p} 
\end{equation}
where the plus sign denotes a large rapidity gap. However the cross section
is expected to be rather small~\cite{9,15}, and as a consequence, the
corresponding event rate appears to be too low at the Tevatron. Only at the
LHC is there a chance of observing this exclusive Higgs production
process~\cite{11,15,16,dkmoz}. Various effects cause a drastic reduction of
the cross section for process~(\ref{excl}), for details see~\cite{9,17}.
First, the proton form factors strongly limit the available phase space in
the transverse momentum of the produced Higgs, $q_{T}\sim 1/R_{p}$, where
$R_{p}$ is the proton radius. Secondly, we have to account for the
probability $\hat{S}^2$ that the gaps survive the soft rescattering effects
of spectator partons which may populate the gaps with secondary particles,
see for example~\cite{5,18,19}.  Thirdly, the cross section is also
suppressed by QCD Sudakov-like radiative effects~\cite{8,9,20}.\par
The cross section is larger in the semi-inclusive case when the
protons may dissociate,
\begin{equation}
\label{semi_inc} 
p\overset{(-)}{p}\rightarrow X + H + Y 
\end{equation}
but the Higgs is still isolated by rapidity gaps. In this case there is no
proton form factor suppression and the QCD ``radiation damage" becomes
weaker. Moreover, a significant contribution to process~(\ref{semi_inc})
comes from Higgs production via $WW/ZZ$ fusion, i.e. $qq\to qqH$. Since this
process is mediated by colourless $t$-channel $W/Z$ exchanges there is no
corresponding gluon bremsstrahlung in the central region~\cite{3,4,5,7}, and
thus Sudakov suppression of the rapidity gaps does not occur. Another
characteristic feature of the vector boson fusion Higgs production process is
that it is accompanied by energetic quark jets in the forward and backward
directions. Recently, interest in this type of Higgs production process at
the LHC has risen rapidly, see for example~\cite{17,18,zknr,28}. The
particular importance of the electroweak fusion process is that it allows a
determination of the Higgs coupling to vector bosons. It is worthwhile to
note that the $WW/ZZ$ fusion mechanism can provide a potential way to
identify $H\rightarrow b\bar{b}$ decays at the LHC, if particular kinematic
configurations with large rapidity gaps are selected, see for
example~\cite{17}.\par The most delicate issue in calculating the cross
section for processes with rapidity gaps concerns the soft survival
factor\footnote{Recall that this factor is not universal, but is very
  sensitive to the spatial distribution of partons inside the colliding
  protons, which in turn results in the dependence on the particular hard
  subprocess as well as on the kinematical configurations of the parent
  reaction~\cite{18,19}.} $\hat{S}^2$. This factor has been calculated in a
number of models for various rapidity gap processes, see for
example~\cite{18,19,21,22}. Although there is reasonable agreement between
these model expectations, it is always difficult to guarantee the precision
of predictions which rely on soft physics.\par Fortunately, the calculations
of $\hat{S}^2$ can be checked experimentally by computing the event rate for
a suitable calibrating reaction and comparing with the observed rate. As
shown in~\cite{9,20} the appropriate monitoring process for the
double-diffractive mechanism is central dijet production with a rapidity gap
on either side. To date, such a check has been the prediction of diffractive
dijet production at the Tevatron in terms of the diffractive structure
functions measured at HERA~\cite{23}. The evaluation of the survival factor
$\hat{S}^2$ based on the formalism of~\cite{18,19} appears to be in
remarkable agreement with the CDF data (see also~\cite{24,25}). We expect
that future measurements in run II of the Tevatron will provide us with
further detailed information on $\hat{S}^2$.\par As was pointed out
in~\cite{26,27}, the survival factor for the gaps surrounding $WW\rightarrow
H$ fusion can be monitored experimentally by observing the closely related
central production of a $Z/W$ boson with the same rapidity gap and jet
configuration. The discussion in~\cite{26,27} concerns the studies of $Z/W+2$
forward jet production with subsequent leptonic $Z$ decay in association with
a rapidity gap trigger, which could allow a substantial suppression of the
QCD-induced backgrounds. It is worthwhile to keep in mind that in different
papers different criteria are used for the definition of the rapidity gaps.
For example, in the approach of~\cite{28,27} no jets with $p_{T}>10-20$~GeV
are permitted within the gaps, while in~\cite{8,9,11} the gap is required to
be completely devoid of any soft hadrons.\par Note that the determination of
the gap survival factor in the vector boson mediated process is interesting
in its own right, since here we can separate the contribution of the short
transverse size component of the proton~\cite{19}, which so far has not
attracted much attention, either theoretically or experimentally.\par The
reader should be warned about the potential problems with the identification
of rapidity gaps in the real life experimental environment at the LHC. When
the LHC operates at medium and high luminosity, the recorded events will be
plagued by overlap interactions in the same bunch crossing (pile-up).
However, as discussed in~\cite{dkmoz}, at least at the medium luminosity of
$10^{33}\textrm{ cm}^{-2}\textrm{s}^{-1}$ the gaps should be detectable.
Using vertex reconstruction information, one can separate particles
originating from the same vertex as the high-$E_{T}$ jets from those
relatively low-$p_T$ particles which arise from other vertices corresponding
to the pile-up interactions. However, it is quite unlikely that this
technique can be used for the super-LHC luminosity of $10^{35}\textrm{
  cm}^{-2}\textrm{s}^{-1}$.\par In this paper we develop the ideas
of~\cite{26,27} further by considering the decays of both (light) Higgs and
$Z$ bosons into $b \bar b$ pairs, the dominant decay channel of the
former\footnote{For completeness, we also consider the $W+2$~jet production
  process.}. In each case we require two forward energetic jets, and rapidity
gaps on either side of the centrally produced decay products. Both $H$ and
$Z$ can be produced by electroweak vector boson fusion, for which gaps are
`natural', but the $Z$ can also be produced via $\mathscr{O}(\alpha_{S}^{2})$
QCD processes, with both quarks and gluons exchanged in the $t$-channel, but
as we shall see these have a smaller soft survival factor. Finally, there is
a potentially large continuum $b \bar b$ background, which is again heavily
suppressed when rapidity gaps are required.\par The paper is organised as
follows. In Section~\ref{sec:partoncalc} we discuss the calculation of the
various signal and background processes at the parton level. In
Section~\ref{sec:partonchar} we perform detailed numerical calculations with
realistic experimental cuts to determine the corresponding cross sections. In
Section~\ref{sec:gap} we explore the consequences of hadronisation and
estimate the gap survival probabilities, $\hat{S}^{2}$, for the various
processes under consideration. Finally, in Section~\ref{sec:results} we
combine the parton-level cross sections with the gap survival probabilities
to give our final predictions for the cross sections. Section~\ref{sec:conc}
summarises our conclusions.

\section{Parton Level Calculation of Higgs, $Z$ and $W$ production}
\label{sec:partoncalc}

In this section we assemble the various parton-level cross sections that are
used in our analysis. We are particularly interested in the overall event
rates for the various signal and background processes, and in the kinematic
distributions of the final-state particles. All matrix elements used in the
cross-section calculations are obtained using MADGRAPH~\cite{Stelzer:1994ta}.
In all cases we work in the zero width approximation for the
centrally-produced bosons.\par
\begin{figure}[ht]
\begin{center}
\scalebox{1}{\includegraphics{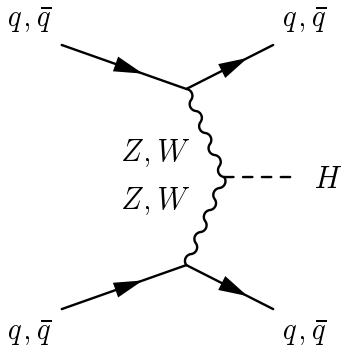}}
\caption{Higgs production via electroweak vector boson fusion.}\label{h}
\end{center}
\end{figure}

 We begin by considering the fundamental signal
process, $\mathscr{O}(\alpha_{W}^{3})$ Higgs production by $WW$, $ZZ$ fusion:
$qq\rightarrow qqH$ (Fig.~\ref{h}).  We assume that the Higgs is light, so
that the dominant decay is into the $b \bar b $ final state. Because the
momentum transfer is much smaller than the energy of the struck quark
jets($\langle p_{T}\rangle \sim M_{W/Z}$), the jets are produced
predominantly at small angle (i.e. large rapidity). Note that there is no
exchange of colour in the $t$-channel, which leads to a suppression of
hadronic radiation in the central region between the forward
jets~\cite{3,4,5,7}.\par
\begin{figure}[ht]
\begin{center}
\scalebox{1}{\includegraphics{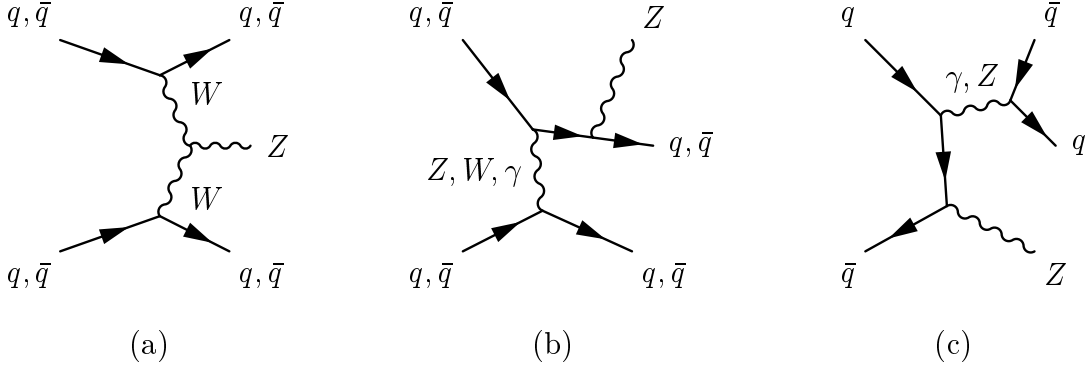}}
\caption{The three topologies for $Zqq$ production via
electroweak vector boson exchanges.}\label{zew}
\end{center}
\end{figure}

 Representative Feynman diagrams for the analogous
$\mathscr{O}(\alpha_{W}^{3})$ $Z$ production process, $q q \rightarrow q q Z$
and $q \bar{q}\rightarrow q \bar{q} Z$ are shown in Fig.~\ref{zew}. They were
first analysed in~\cite{26,27}.  Note that in addition to the $WW$ fusion
diagram, Fig.~\ref{zew}(a), the $Z$ can also be radiated off either of the
incoming or outgoing quark lines, Figs.~\ref{zew}(b) and (c). The
characteristic topology of (b) is of a $Z$ preferentially produced in the
forward or backward region close in rapidity to one of the final-state quark
jets. Requiring {\it centrally} produced $Z$ decay products tends to suppress
this contribution. Process (c) corresponds to $s$-channel production of the
final-state $q \bar q$ pair, with the $Z$ boson emitted off the incoming
quark lines. It does not correspond to $t$-channel colour singlet exchange
and is heavily kinematically suppressed by requiring a large rapidity
separation between the jets.\par
\begin{figure}[ht]
\begin{center}
\scalebox{1}{\includegraphics{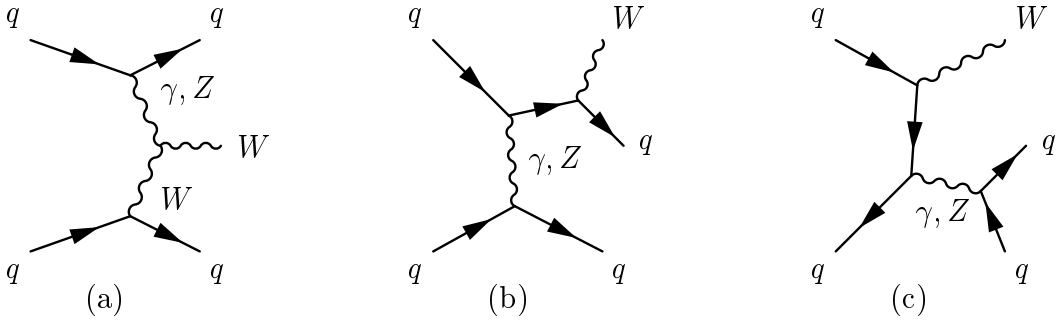}}
\caption{The three topologies for $W$ production.}\label{qqw}
\end{center}
\end{figure}

 Similar remarks apply to $W$ production.
Representative Feynman diagrams for $qq \to Wqq$ are shown in Fig.~\ref{qqw}.
Note that the central $W$-production via $\gamma$ exchange corresponding to
Fig.~\ref{qqw}a was recently discussed in~\cite{kmrph}.

\begin{figure}[ht]
\begin{center}
\scalebox{1}{\includegraphics{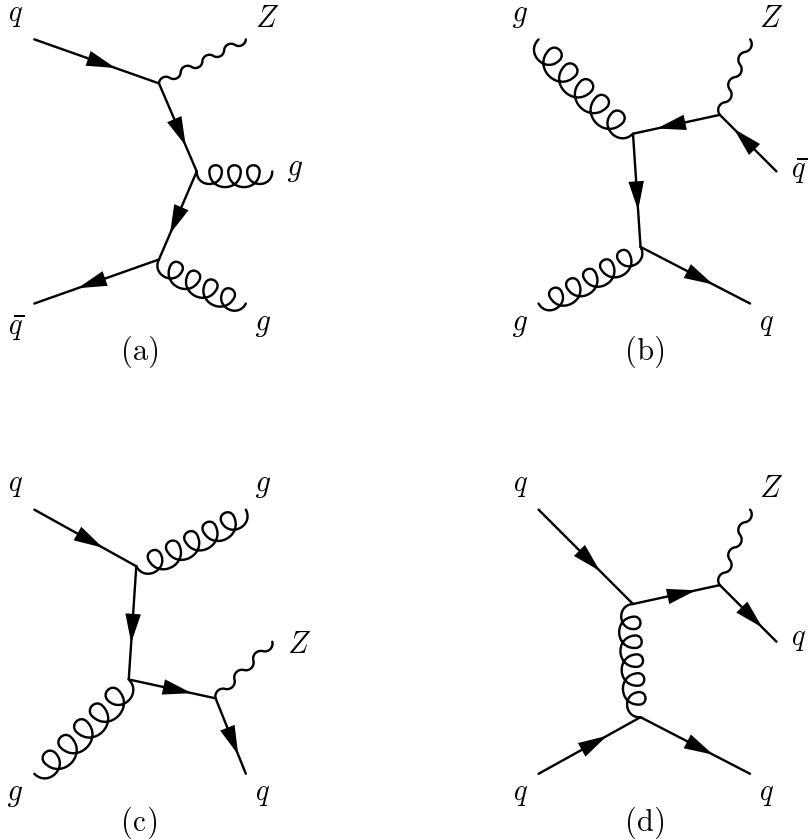}}
\caption{QCD background processes to $Z$ production.}\label{zqcd}
\end{center}
\end{figure}

The above $\mathscr{O}(\alpha_{W}^{3})$ $H$ and $Z$ production processes both
therefore give rise to rapidity gap signatures between the forward jets and
the central $H$ and $Z$ decay products. However there is a potentially
important QCD $\mathscr{O}(\alpha_{S}^{2}\alpha_{W})$ background contribution
to $Z\; +\; $2~jet production where the internal electroweak gauge boson is
replaced by a {\it gluon}.  More generally, at this order indistinguishable
background contributions can arise from any $2 \to 2$ scattering process
(other than $gg \to gg$) where the $Z$ is radiated off a quark line.
Representative Feynman diagrams are shown in Fig.~\ref{zqcd}. By selecting
forward jets and central $Z$ bosons, in order to mimic the dominant Higgs
configuration, the $t$-channel momentum transfer is minimised, and these QCD
processes split into two types: $t$-channel quark (Figs.~\ref{zqcd}(a,b,c))
and gluon exchange (Figs.~\ref{zqcd}(d)). Requiring rapidity gaps therefore
suppresses both type of contribution, as will be discussed in
Section~\ref{sec:gap} below.

\begin{figure}[ht]
\begin{center}
\scalebox{1}{\includegraphics{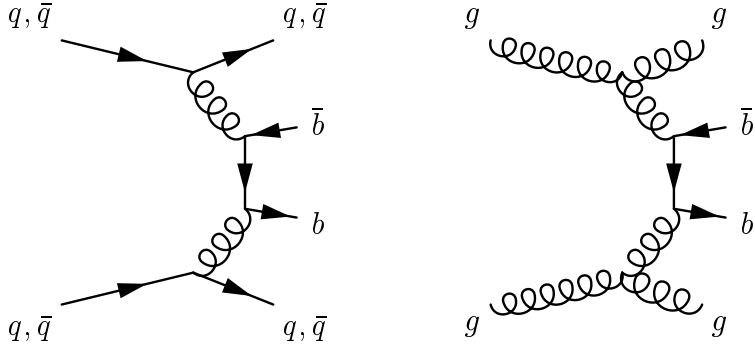}}
\caption{QCD backgrounds to $qq\rightarrow qq(H,Z),\ (H,Z)\rightarrow
b\bar{b}$.}\label{hbk}
\end{center}
\end{figure}

Finally, given that we are interested in the $b\bar b$ decay modes of both
the Higgs and $Z$ bosons, with two additional jets in the final state, there
is a class of $\mathscr{O}(\alpha_{S}^{4})$ pure-QCD background processes of
the form $ab \to cd +b \bar b$ with $a...d = q,g$, examples of which are
shown in Fig.~\ref{hbk}. We will consider the corresponding cross sections in
the following section, with the additional requirement that $m_{b\bar b}
\simeq M_Z$.

\section{Signal and Background Rates and Properties}
\label{sec:partonchar}
In this section we calculate cross sections and kinematic distributions for
the processes described in the previous section, for proton-proton collisions
at the LHC using a representative set of cuts on the final state particles.
We have in mind final states with a jet registered in a forward detector with
pseudorapidity $\eta_{1} > \eta_{\textrm{min}}$, another produced backwards
with $\eta_{2} < -\eta_{\textrm{min}}$, and the $H$, $Z$ and $W$ decay
products produced centrally, with rapidity $\vert y_{H,Z,W} \vert <
y_{\textrm{max}}$. We impose a minimum transverse momentum cut $\ptmin$ on
all final-state jets.

\subsection{Total cross sections}
\label{subsec:total}

\begin{figure}
\hspace{15mm}
\scalebox{0.45}{\includegraphics[angle=-90]
{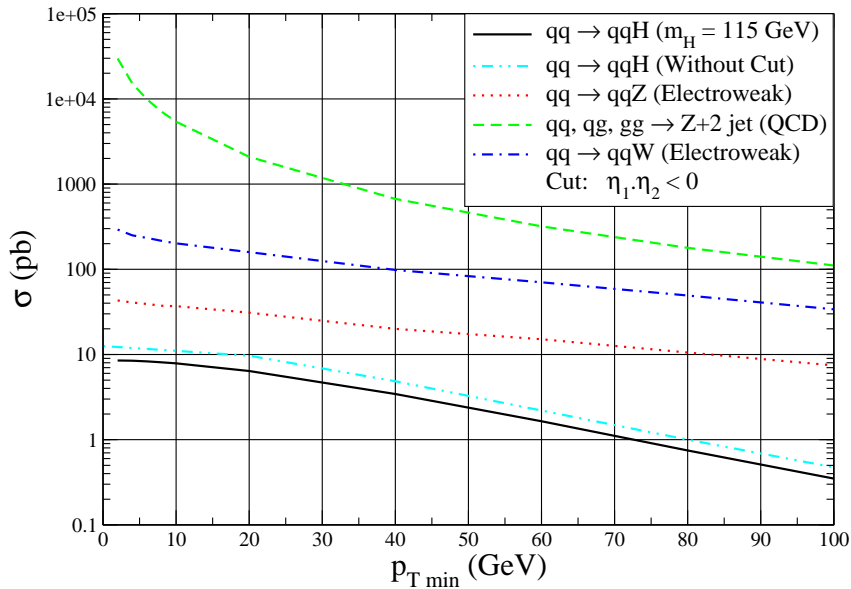}}
\caption{Total cross sections for ($H,\; Z, \; W$)~$+2$~jet production in $pp$ collisions at
$\sqrt{s}=14$ TeV as a function of the cut on the jet transverse momentum, $\ptmin$.
Rapidity cuts on the final state jets are also imposed, as indicated.}
\label{totx}
\end{figure}

Figure~\ref{totx} shows the total cross section for Higgs, electroweak $Z$
and $W$, and QCD $Z$ production (with no branching ratios included) as a
function of a cut on the minimum transverse jet momentum $\ptmin$. The Higgs
mass is $M_H = 115$~GeV and the leading-order MRST98LO \cite{Martin:1998sq}
parton distribution set is used. Note that only for $H$ production is the
cross section finite in the limit $\ptmin \to 0$\footnote{The possibility of
  exchanging a massless photon or gluon in the $t$ channel gives rise to an
  infrared singularity in the electroweak and QCD $Z\; +\; $2~jet production
  processes as $\ptmin \to 0$, see Figs.~\ref{zew} and \ref{zqcd}. For
  exclusive $pp\rightarrow p + X + p$ collisions this singularity is cut off
  by the $t_{\textrm{min}}$-effect.}. In addition, the possibility that the
final state jets in $Z$ and $W$ production originate in the splitting process
$g^*, \gamma^* \to q \bar q$ (for example, see Fig.~\ref{zew}(c)) requires a
jet separation cut. The minimal way to do this is simply to require that one
of the jets is produced in the forward hemisphere and the other in the
backward hemisphere, i.e. $\eta_1\cdot\eta_2 < 0$. When we come to consider
`realistic' cuts, in particular to isolate the jets from each other and the
$H$ and $Z$ decay products, we will impose a large rapidity separation cut in
which one jet is produced far forward and one far backward: $\vert
\eta_1\vert, \; \vert \eta_2 \vert > \eta_{\textrm{min}}$, $\eta_1\cdot\eta_2
< 0$. For the Higgs production process, which has no infrared or collinear
singularities, the imposition of $\ptmin$ and $ \eta_1\cdot\eta_2 < 0$
acceptance cuts simply reduces the cross section slightly (by approximately
25\% for a broad range of $\ptmin$ values), see Fig.~\ref{totx}.\par
\begin{figure}
\hspace{25mm}
\scalebox{0.4}{\includegraphics[angle=-90]
{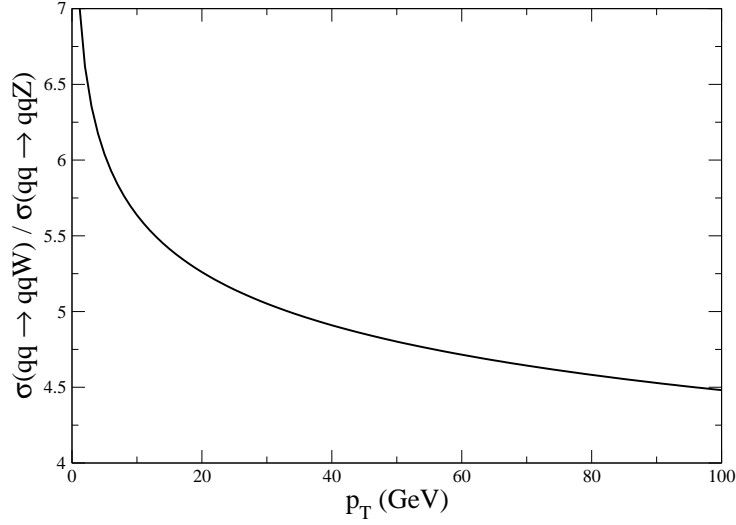}}
\caption{The ratio between the total cross sections of the $W$ and $Z$
(electroweak) production processes as defined in the previous figure. The $W$
cross section is `more divergent' than that for  $Z$ production at low $p_{T}$, because of the
extra photon-exchange diagram involving the triple gauge boson vertex (Fig.~\protect\ref{qqw}(a)).}
\label{wzratio}
\end{figure}
\begin{figure}
\hspace{25mm}
\scalebox{0.4}{\includegraphics[angle=-90]
{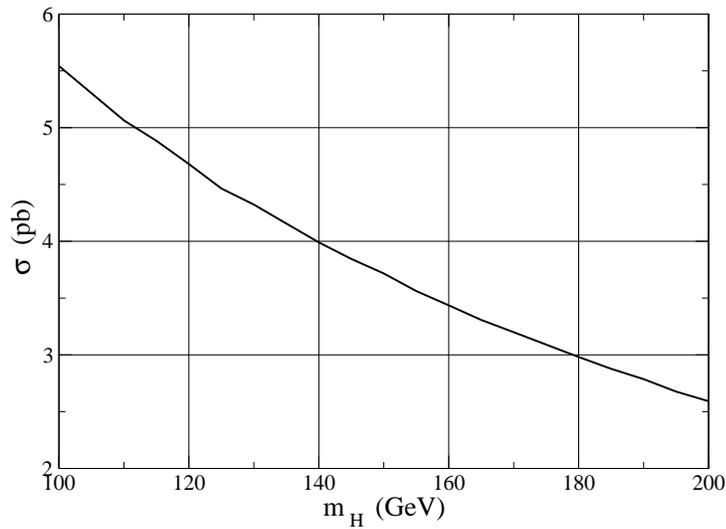}}
\caption{The $\sqrt{s} = 14$ TeV $qq\rightarrow qqH$ cross section as a function of the Higgs mass $M_H$.}
\label{mhvar}
\end{figure}

Figure~\ref{totx} shows that there is a strong ordering of the cross sections
$\sigma(Z, QCD) \gg \sigma(Z,EW) \gg \sigma(H)$, with $\sigma(Z, QCD)$
exhibiting the strongest dependence on $\ptmin$. The $W$ cross section has a
stronger infra-red singularity as $\ptmin\to 0$ than the corresponding $Z$
cross section, due to the soft photon singularity present in the extra
diagram with respect to the $Z$ production process involving the triple gauge
boson vertex (Fig.~\ref{qqw}(a)). This is shown more clearly in $W/Z$ cross
section ratio plot, Fig.~\ref{wzratio}. The Higgs cross section is only
weakly dependent on the mass $M_H$, decreasing by a factor of 2 as $M_H$
increases from 100~GeV to 200~GeV, see Fig.~\ref{mhvar}.\par
Note that all the above cross sections are evaluated in the zero $Z/W$ width
approximation and at leading order in perturbation theory. In particular, in
the QCD $Z+2$~jet calculation the scale of the strong coupling $\alpha_S$ is
not determined, and there is a non-negligible scale dependence uncertainty as
a result. We use $\alpha_S\equiv \alpha_S(M_Z^2)$ throughout. One could also
argue for a smaller scale characteristic of the transverse momenta of the
forward jets, e.g. $\alpha_S \equiv \alpha_S(\ptmin^2)$. We will discuss the
impact of such a choice on our predicted event rates in
Section~\ref{sec:results}.


\subsection{Distributions}
\label{subsec:dist}

\begin{figure}
\hspace{38mm}
\includegraphics[scale=0.5]
{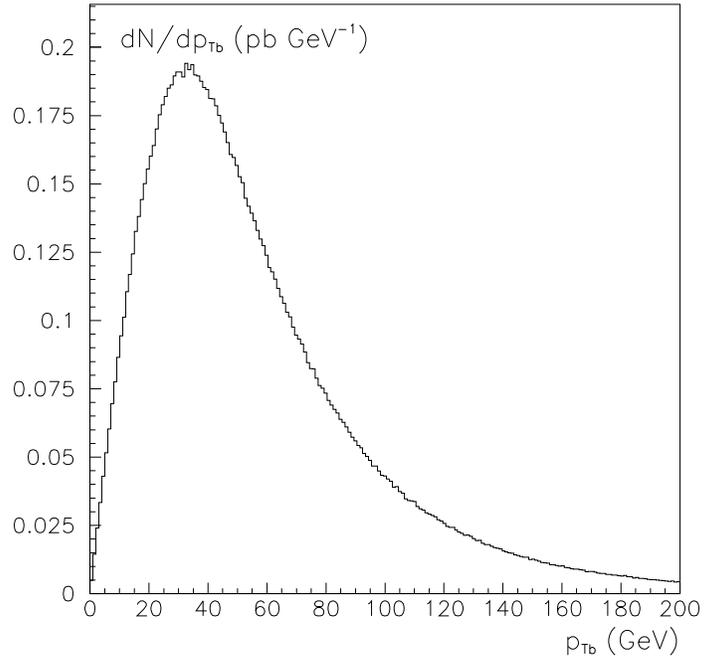}
\caption{Jet transverse momentum distribution for $qq\rightarrow qqH$. The peak is at around $M_{W}/2$.}\label{ptp}
\end{figure}

\begin{figure}
\hspace{38mm}
\includegraphics[scale=0.5]
{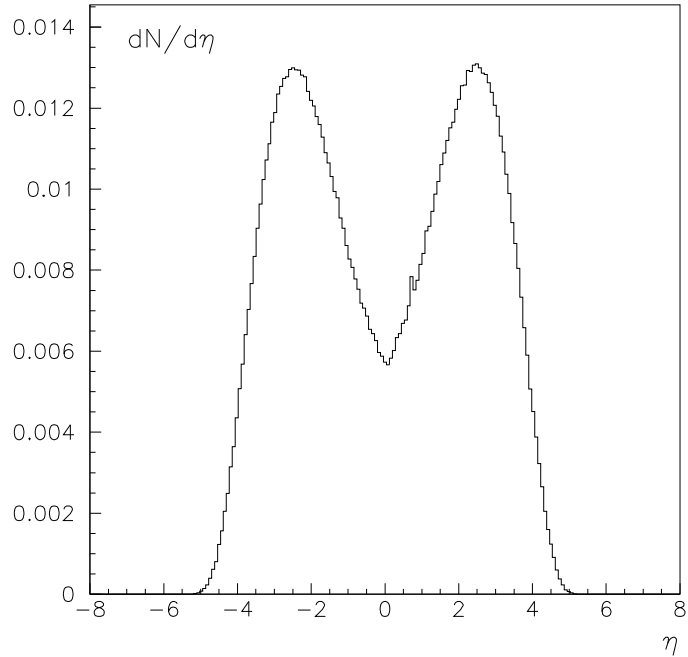}
\caption{Rapidity distribution of jets for $qq\rightarrow qqH$ with $p_{T min}=40$GeV and $\eta_{1}.\eta_{2}<0$. The vertical scale is normalised arbitrarily.}\label{rap_h}
\end{figure}

\begin{figure}
\hspace{40mm}
\includegraphics[scale=0.5]
{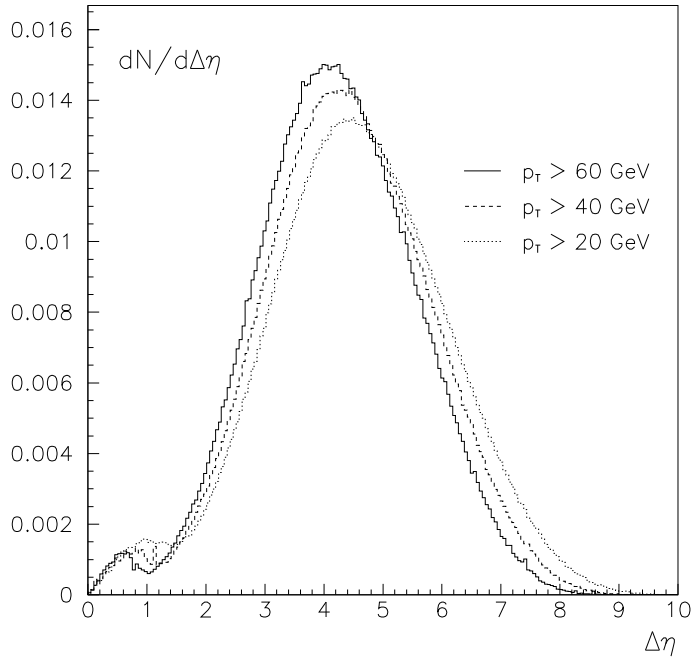}
\caption{Dijet rapidity difference for $qq\rightarrow qqH$ (with $\eta_{1}.\eta_{2}<0)$ as a function of the $\ptmin$ transverse momentum cut. The vertical scale is normalised arbitrarily.
The gap narrows marginally as the cut is raised, as expected from kinematics. The small excess seen at low $\Delta \eta$ is discussed in the text.}
\label{vrap_h}
\end{figure}

\begin{figure}[ht]
\begin{center}
\scalebox{1}{\includegraphics
{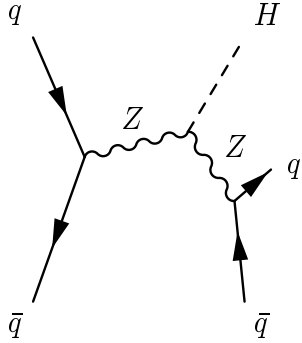}}
\caption{Contribution to the $\mathscr{O}(\alpha_{W}^{3})$ electroweak process
$q \bar q \to H q \bar q$ that resonates when $m_{jj} \sim M_Z$.}\label{zres}
\end{center}
\end{figure}

\begin{figure}
\hspace{40mm}
\includegraphics[scale=0.5]
{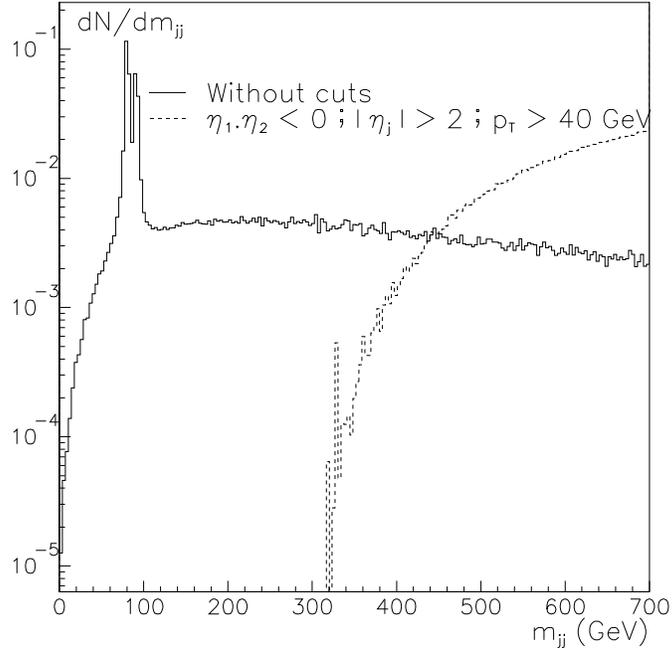}
\caption{The dijet invariant mass for $qq\rightarrow qqH$ shows a double
resonance around $m_{Z}$ and $m_{W}$ -- due to diagrams such as Fig.~\protect\ref{zres}. The vertical scale is normalised arbitrarily. When the jets are required to be separated and forward in rapidity the effect is irrelevant.}
\label{mjj}
\end{figure}

Our objective is to find a set of selection cuts that minimises the
background while not affecting drastically the Higgs, $Z$ and $W$ rapidity
gap signal.  We begin by calculating the transverse momentum and rapidity
distributions of the jets in $qqH$ production, Figs.~\ref{ptp} and
\ref{rap_h}. Evidently the jets are predominantly produced with transverse
momenta of order $M_W/2 \sim 40$~GeV, and with a rapidity separation of
around 5, see Fig.~\ref{vrap_h}.  Notice the small excess around $\Delta \eta
\sim 1/2$. This is caused by the contributing process $q \bar q \to H q \bar
q$ in which $m_{jj} \sim M_Z$, i.e. the Higgs is produced in association with
a $Z$ (or $W$) boson which subsequently decays into a $q \bar q$ pair, see
Fig.~\ref{zres}. This is more clearly seen in the dijet mass distribution,
Fig.~\ref{mjj}.

Requiring the jets to be well-separated in rapidity forces $m_{jj}$ to be
large and this resonant contribution is strongly suppressed. For example,
Fig.~\ref{mjj} also shows the dijet mass distribution for $\vert \eta_{1,2}
\vert > \eta_{\textrm{min}} = 2$.\par
\begin{figure}
\hspace{38mm}
\includegraphics[scale=0.5]
{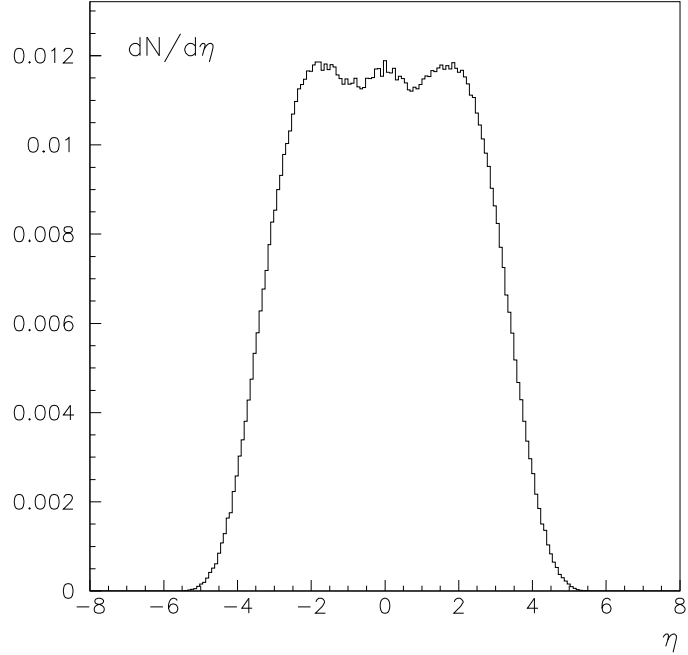}
\caption{Rapidity distribution of jets for electroweak $Z$ production. The vertical scale is normalised arbitrarily. The gap
is narrower than for the Higgs signal (Fig.~\ref{rap_h}). A jet transverse
momentum  cut of $\ptmin = 40$~GeV is applied, as is $\eta_{1}.\eta_{2}<0$.}
\label{rap_zew}
\end{figure}

 The jet rapidity distribution for
electroweak $qqZ$ production is shown in Fig.~\ref{rap_zew}.  Comparing with
Fig.~\ref{rap_h} for $qqH$, we see that the jets produced with a $Z$ are more
uniform in rapidity, The `$WW$-fusion' diagrams of Fig.~\ref{zew}(a) still
produce jets with a large separation, but the central region is now filled in
by contributions from the other non-fusion `$Z$-bremsstrahlung' processes,
Figs.~\ref{zew}(b,c).  Electroweak $W$ production has very similar
characteristics to electroweak $Z$ production.\par
\begin{figure}
\hspace{38mm}
\includegraphics[scale=0.5]
{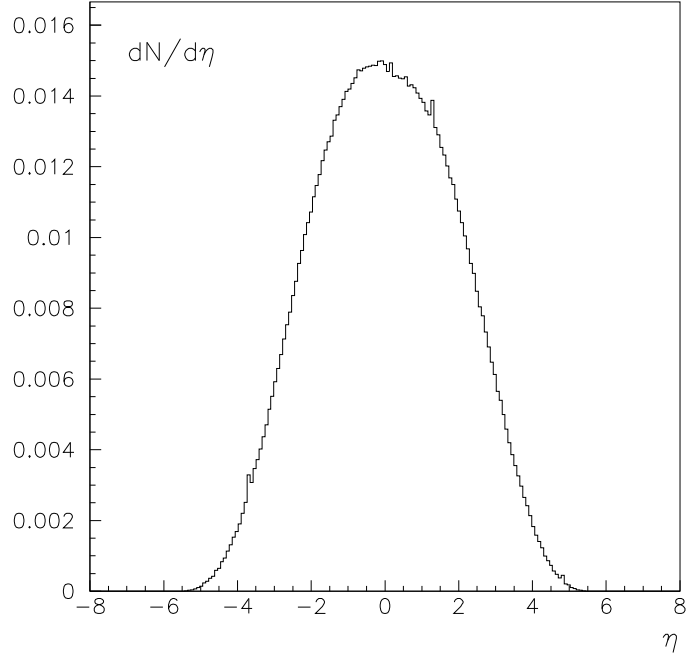}
\caption{Rapidity distribution of jets for QCD $Z\; +\; 2$~jet production. The vertical scale is normalised arbitrarily. A jet transverse
momentum  cut of $\ptmin = 40$~GeV is applied, as is $\eta_{1}.\eta_{2}<0$.}
\label{rap_zqcd}
\end{figure}

\begin{figure}
\hspace{40mm}
\includegraphics[scale=0.5]
{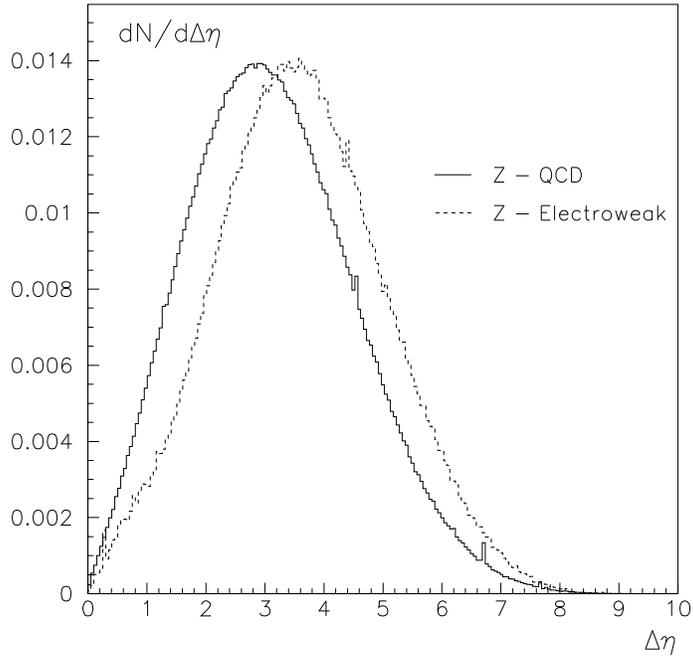}
\caption{Comparison of the dijet rapidity differences for the electroweak and
QCD $Z$ production processes. The vertical scale is normalised arbitrarily. A jet transverse
momentum  cut of $\ptmin = 40$~GeV is applied, as is $\eta_{1}.\eta_{2}<0$.}
\label{zeppencomp}
\end{figure}

 For the QCD background to
electroweak $Z$ production, the jets are produced much more centrally, see
Fig.~\ref{rap_zqcd}. Requiring a jet in each forward/backward hemisphere
leads to a typical rapidity separation of about 3, as shown in
Fig.~\ref{zeppencomp}, which is significantly less than for either $H$ or
electroweak $Z$ production. There is no natural rapidity gap, as for the
$t$-channel colour-singlet exchange processes.

\subsection{Selection cuts}
\label{subsec:sel}

We can now proceed to define a set of selection cuts that leads to a sample
of $H$, $Z$ and $W$ events with the potential to exhibit rapidity gaps.
Since our primary goal is to calibrate the gap survival for Higgs production,
we will concentrate first on the $b\bar b$ decays of $H$ and $Z$, the latter
produced either via electroweak or QCD processes.\par When considering the
$b\bar b$ decay modes of both the Higgs and $Z$ bosons, we must include also
the important irreducible background from QCD $\mathscr{O}(\alpha_{S}^{4})$
$b\bar b\; +\; $2~jet production, see Fig.~\ref{hbk}.  Such processes give a
continuous distribution of $m_{b\bar b}$ masses, and in what follows we
impose a cut of $\vert m_{b\bar b} - M_Z \vert < 10$~GeV to select those
background events that mimic $Z\to b \bar b$ decay.\par The configuration we
have in mind has one jet registered in a forward detector with $\eta >
\eta_{\textrm{min}}$, another produced backwards with $\eta <
-\eta_{\textrm{min}}$, and the two $b$ jets from $H$ and $Z$ decay produced
centrally.  From the results of the previous section, such a selection will
in principle preserve the bulk of the Higgs signal while suppressing the
(non-gap) QCD $Z$ and $b\bar b\; +\; $2~jet production.\par For both ATLAS
and CMS, the forward hadron calorimeters cover approximately $3 < \vert \eta
\vert < 5$, and so we will require our forward dijets to be produced in this
region of rapidity, with $p_T > \ptmin = 40$~GeV. In order to separate the
$H,\; Z$ decay jets from the forward jets, we require $\vert \eta_b\vert <
1.5$, and $p_{Tb} > 10$~GeV\footnote{The typical transverse momentum of the
  jets in both the signal and background processes is $\sim M_Z/2$, and this
  cut does not have any significant effect on the event rates, see for
  example Table~\ref{cuts} below.}. Although these cuts are designed to
reflect the `natural' characteristics of $qqH$ production, they do result in
a non-negligible loss of signal rate, even before $b$-tagging efficiencies
etc. are taken into account. This is illustrated in Table~\ref{cuts}, which
quantifies the effect on the cross section of applying the cuts sequentially.
One can see that imposing forward jet cuts has the largest impact on the
cross section, and indeed this is the case for all the processes considered.

\begin{table}[!h]
\begin{center}
\begin{tabular}{|c|c|c|}
\hline
Cut Imposed & Cross Section for $qq\rightarrow qqH$ at $p_{T} > 40$GeV &
 \% of Initial Cross Section \\
\hline\hline
  & $4.86$pb & 100\% \\
\hline
Br($H\rightarrow b\bar{b}$) & 3.49pb & 71.9\% \\ \hline
$\eta_{1}\cdot\eta_{2} < 0$ & 2.47pb & 50.8\% \\ \hline
$\Delta \eta_j > 6$ & 0.495pb & 10.2\% \\ \hline
$|\eta_{j}| > 3$ & 0.0990pb & 2.04\% \\ \hline
$|\eta_{b}| < 1.5$ & 0.0465pb & 0.957\% \\ \hline
$p_{Tb} > 10$~GeV & 0.0463pb & 0.953\% \\ \hline
\end{tabular}
\caption{Loss of $qq\rightarrow qqH$ cross section at $\sqrt{s} = 14$~TeV with $M_{H}=115$~GeV in applying selection cuts and the $b \bar b$ branching ratio.}
\label{cuts}
\end{center}
\end{table}

\begin{figure}
\hspace{15mm}
\scalebox{0.45}{\includegraphics[angle=-90]
{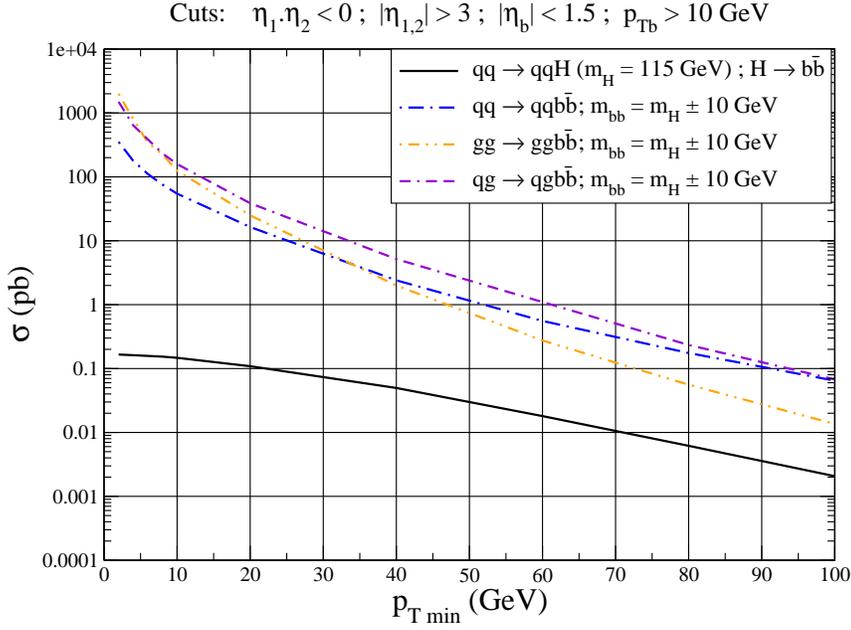}}
\caption{Cross sections at $\sqrt{s} = 14$ TeV for Higgs production processes
after application of the cuts described in the text.}
\label{cut_sig_H}
\end{figure}

\begin{figure}
\hspace{15mm}
\scalebox{0.45}{\includegraphics[angle=-90]
{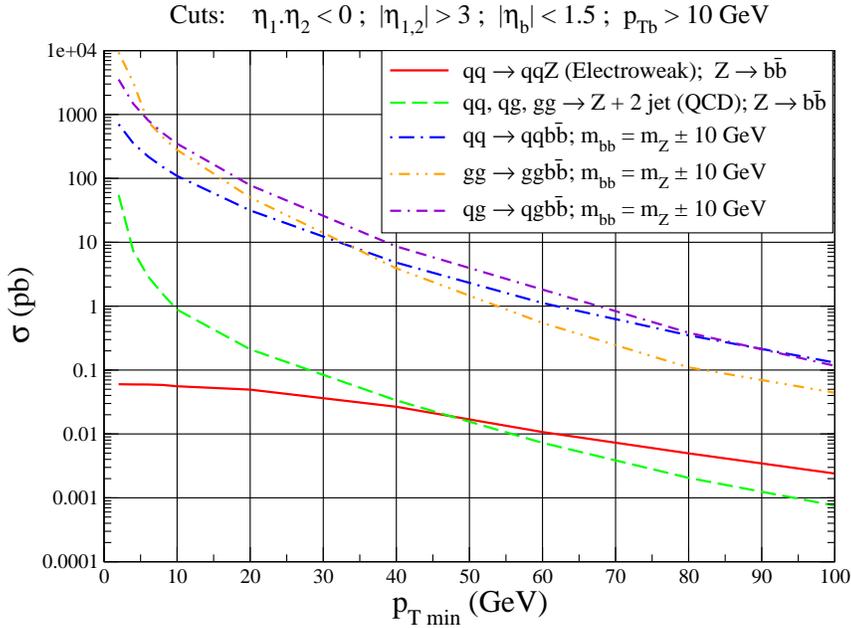}}
\caption{Cross sections at $\sqrt{s} = 14$ TeV for $Z$ production processes
after application of the cuts described in the text.}
\label{cut_sig_Z}
\end{figure}

Figures \ref{cut_sig_H} and \ref{cut_sig_Z} show the cross sections at
$\sqrt{s}=14$ TeV as a function of $\ptmin$ for all processes. The Higgs
production cross section is reduced by a factor of $\sim 100$ and the
electroweak $Z$ production by $\sim 1000$ in comparison with
Figure~\ref{totx}\footnote{This is because of the difference in rapidity
  distributions in the $H$ (\Figs{rap_h}{vrap_h}) and $Z$
  (\Figs{rap_zew}{zeppencomp}) cases which is caused by the process shown in
  Figs~\ref{zew}(b),(c) and \Fig{zqcd}(d) where the quark jets are closer to
  each other.}. The cuts reduce the $Z$ production QCD background by a factor
of $\sim 10000$. As already mentioned, in evaluating the pure QCD $b\bar{b}$
production cross sections we further impose the restriction that the dijet
invariant mass be within $10$ GeV of $M_{Z}$.

\section{Gap Survival Probability}
\label{sec:gap}
\subsection{Parton Level}
In the previous sections we have presented cross sections for Higgs, Z and
QCD $b\bar{b}$ production processes for events with rapidity gaps at the
parton level. We take the definition of rapidity gap to mean that there
should be no minijets with a large ($p_{T}>10$~GeV) transverse momentum
within the gap region. As discussed in the introduction, the selection of
rapidity gap events improves the signal to background ratio because gaps are
a characteristic feature of the vector boson fusion process, whereas they are
not for QCD $Z$ and continuum $b\bar{b}$ production. Of course, the results
presented in the previous section should be corrected to account for the
rescattering of spectator partons, that is the possibility that another pair
of initial, fast partons interacts inelastically in the same event.
Secondaries produced in this inelastic interaction may fill the gap and the
probability, $\hat{S}^2$, for the gap to survive depends on the criteria used
to select the gap. Insofar as we require only to have no high $p_{T}$
particles (or minijets) within the gap interval, the effect is not too large.
The probability to produce high $p_{T}$ secondaries is relatively small and
the corresponding survival factor $\hat{S}^2\simeq 0.7-0.9$ (depending on the
$p_{T}$ cut) is more or less close to one, see for example~\cite{28}. For any
specific kinematics (and $p_{T}$ cuts), the value of $\hat{S}^2$ for such
`parton level' gaps may be estimated using, for example, one of the options
of the PYTHIA Monte Carlo program~\cite{pythia}, or an `exponentiation' model
such as that used in~\cite{RZ}\footnote{ Note that the large survival
  probability $\hat{S}^2\sim 0.8$ used in~\cite{28} corresponds just to
  parton level gaps, and was calculated using the model of~\cite{RZ}.}.

\subsection{Hadron Level}

\begin{figure}[ht]
\begin{center}
\scalebox{1}{\includegraphics
{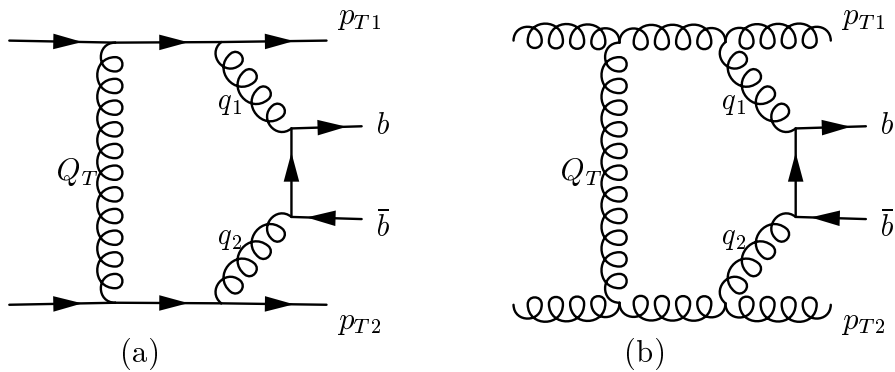}}
\end{center}
\caption{Screening of QCD dijet + $b\bar{b}$ production via gluon exchange.}\label{screen}
\end{figure}

As seen in Figs.~\ref{cut_sig_H},~\ref{cut_sig_Z}, the QCD-induced $b\bar{b}$
background is still large. It exceeds by two orders of magnitude the $Z/H$
cross sections and it is therefore necessary to suppress the background
further. This can be done by requiring a {\it completely clean} gap, i.e.
without any soft hadrons. Indeed, all the QCD processes we consider are
characterised by gluon (or quark) $t$-channel exchange, which unavoidably
produces a colour flow along the gap. During hadronisation this colour flow,
in turn, creates plenty of soft secondaries which fill the gap. On the other
hand, there is no such effect for the electroweak graphs (Figs.~\ref{h},
~\ref{zew}a, \ref{zew}b) since the vector boson exchange is colourless. This
means that if we require clean hadron-level gaps we can immediately discard
the diagrams of Fig.~\ref{hbk}. The only way to create a gap in a QCD induced
event is to screen the colour flow (across the gap) by an additional gluon
(or quark) exchange; that is, to consider graphs of the type shown in
Fig.~\ref{screen}.\par
Note that in leading order we can screen the colour flow in both gaps (above
and below the $b\bar{b}$-pair) with only one additional $t$-channel gluon,
with momentum $Q_{T}$ say. The price we pay for this screening is a factor of
$\alpha_{S}$ supplemented by the $dQ_{T}^2$ loop integration in each
amplitude; that is ($\int\alpha_{S}\dots d^{2}Q_{T}$)$^2$ in the cross
section. At first sight, the major contribution comes from the small $Q_{T}$
region where the QCD coupling $\alpha_{S}$ is larger. Moreover, the integral
takes the form
\begin{equation} 
\int\alpha_{s}\frac{dQ_{T}^2}{Q_{T}^2} 
\end{equation}
and has infrared logarithmic divergence at $Q_{T}\ll p_{T\textrm{jet}1,2}$.
However, this divergence is cut off by the effects of higher order double
logarithmic QCD radiation, see for example~\cite{8,9,20}. The point is that a
small $Q_{T}$ gluon screens the hard gluon at rather large distances $r\sim
1/Q_{T}$ only.  Thus a `hard' gluon $q_{i=1,2}$ may emit a new `semihard'
gluon jet, with transverse energy $E_{T}$ ranging from $Q_{T}$ up to
$q_{iT}=|\vec{p}_{Ti}-\vec{Q}_{T}|$ in the whole rapidity gap interval
$\Delta\eta_{i}$. The leading logarithms come from the $Q_{T}\ll E_{T}\ll
p_{Tjet}$ domain where the expected mean number of these secondary gluons is
\begin{equation}
\label{mean}
\bar{n}_{i}\simeq
\frac{N_{c}\alpha_{S}}{\pi}\Delta\eta_{i}\ln\frac{p_{Ti}^{2}}{Q_{T}^2}. 
\end{equation}
At the amplitude level the corresponding suppression factor describing the
probability for not having such an emission (which otherwise destroys the
gap) has the Sudakov-like form
\begin{equation}
\label{sudakov} \exp(-\bar{n}_{i}/2)=\left(\frac{Q_{T}}{p_{Ti}}\right)^{\frac{N_{c}\alpha_{S}}{2\pi}\Delta\eta}.
\end{equation}
Including this factor in the loop integral, we eliminate the infrared
divergence and obtain the probability, $P_{a}$ ($a=qq,qg,gg$ depending on the
initial state), to screen out the octet (gluon-like) colour flow in $qq$
($gg$ or $qg$) interactions,
\begin{equation}
\label{prob} P_{a}=C_{a}\left(\int^{\ptmin}_{Q_0}\alpha_{S}(Q_{T}^{2})\frac{dQ_{T}^{2}}{Q_{T}^2}\exp\left\{-\frac{N_{c}\Delta\eta}{2\pi}
\int_{Q_{T}}^{\ptmin}\alpha_{S}(Q^{\prime2})\frac{dQ^{\prime2}}{Q^{\prime2}}\right\}\right)^{2}
=C_{a}\left(\frac{2\pi}{N_{c}\Delta\eta}\right)^{2}. 
\end{equation}
Here $\Delta\eta=\Delta\eta_{1}+\Delta\eta_{2}$ is the overall length of the
gaps and, within leading logarithm accuracy, we have put the upper limits in
the $Q_{T}\;(Q^{\prime})$ integration equal to the minimum $p_{T}$ of the
jets. In order to arrive at the right-hand side of~\Eqn{prob} it is
convenient to recast the integral in~\Eqn{prob} as
\begin{equation} \left(\frac{2\pi}{N_{c}\Delta\eta}\right)d\mathcal{J}\exp(-\mathcal{J}(\ptmin,Q_{T})) 
\end{equation}
with 
\begin{equation} \mathcal{J}=\frac{N_{c}\Delta\eta}{2\pi}\int^{\ptmin}_{Q_{T}}\alpha_{s}(Q^{\prime2})\frac{dQ^{\prime2}}{Q^{\prime^2}}.
\end{equation}
Performing the integration we neglect the term
$\exp(-\mathcal{J}(\ptmin,Q_{0}))$ corresponding to the lower limit of
integration. This can always be done safely if we can continue the
perturbative calculation down to the (rather low) scale where the quantity
$\alpha_{s}(Q_0^2)\cdot\Delta\eta$ becomes large. Instead of the conventional
double logarithm expressions (\Eqns{mean}{sudakov}) with a fixed coupling
$\alpha_{S}$, in \Eqn{prob} we have used the running coupling in order to
demonstrate that the result does not depend on whether one accounts for the
running $\alpha_{S}$ or not. The colour factors $C_{a}$ are
\begin{equation}
\label{cfs}
C_{qq}=\frac{C_{F}^2}{(N_{c}^{2}-1)}=\frac{C_{F}}{2N_{c}}=\frac{2}{9},\qquad
C_{gg}=\frac{N_{c}^{2}}{N_{c}^{2}-1}=\frac{9}{8},\qquad
C_{qg}=\frac{C_{F}N_{c}}{N_{c}^{2}-1}=\frac{1}{2}. 
\end{equation}\par
A more precise way to calculate the contributions of \Fig{screen} including
QCD radiative effects is to replace the two gluon $t$-channel exchange by the
non-forward BFKL amplitude~\cite{Lip:86}. For the asymmetric ($Q_{T}\ll
q_{ti}$) configuration the non-forward amplitude contains the double
logarithmic factor of \Eqn{sudakov}, while the single logarithmic
($\sim\mathcal{O}(\alpha_{s}\Delta\eta)$) contribution in this asymmetric
kinematical situation is suppressed, giving a less than $10\%$ correction to
the amplitude (see~\cite{20,RF} for a more detailed discussion). Thus we come
back to the result of \Eqn{prob}. Strictly speaking, besides the suppression
factor \Eqn{sudakov} hidden in the BFKL amplitude, there should be another
Sudakov-like double logarithmic form factor which reflects the absence of QCD
radiation in the interval of gluon transverse momentum between
$p_{T\textrm{jet}}$ and half of the boson (or $b\bar{b}$) mass, $M/2$.
However, in our case the transverse momentum of the jets is
$p_{T}>\ptmin=40$~GeV, which is close to half the boson mass $M_{Z,H}/2$.
Therefore the form factor becomes close to one and we can neglect it.\par
Another point we have to take into account is the fact that now the
$b\bar{b}$-pair may be produced in a colour singlet state only, and the
ordinary $gg\rightarrow b\bar{b}$ hard subprocess cross section (which
includes both colour singlet and octet contributions) should be replaced by
the pure colour singlet cross section~\cite{Khoze:2001xm}
\begin{equation}
\label{singlet} 
\frac{1}{N_{c}^{2}-1}\frac{d\hat{\sigma}^{\rm
incl}}{dt} (gg^{PP}\rightarrow q\bar{q})=\frac{\pi\alpha_S^2}{(N_{c}^{2}-1)E_T^2
M^2}\frac{1}{6} \left[\left(1-\frac{2 E_T^2}{M^2}\right)\left(1-
\frac{2m_q^2}{E_T^2}\right)+\frac{m_q^2}{E_T^2}(1+\beta^2)\right], 
\end{equation} 
where $\beta=\sqrt{1-\frac{4m_q^2}{M^2}}$ and $m_{q}$ is the quark mass. Note
that for the colour singlet production case there is an additional colour
factor $1/(N_{c}^{2}-1)$ which suppresses the QCD background, as the two
colliding gluons are forced to have the same colour.

\subsection{Quark Exchange}

\begin{figure}[ht]
\begin{center}
\scalebox{1}{\includegraphics
{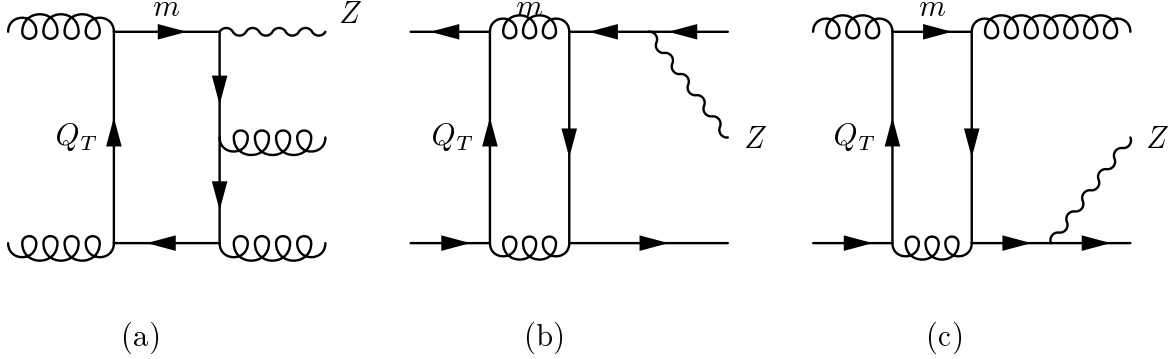}}
\caption{Screening of QCD dijet + $Z$ production via quark exchange.}\label{screenq}
\end{center}
\end{figure}

It is more difficult to screen the colour triplet flow originated by the
quark exchange which we deal with in the electroweak and QCD $Z+2$~jet
processes shown in Figs.~\ref{zew}c, \ref{qqw}c or \ref{zqcd}a,b,c. For
example, to screen the quark colour in Fig.~\ref{zqcd} we have to replace the
graphs Figs.~\ref{zqcd}a,b,c by those of Fig.~\ref{screenq}.
Due to the spin $1/2$ nature of a quark, the large rapidity gaps are
suppressed at the {\it amplitude} level (in comparison with the corresponding
Figs.~\ref{zqcd}a,b,c amplitude contribution) by the factor
$e^{-\Delta\eta/2}$ (i.e. a factor $e^{-\Delta\eta}$ in the cross section).
On the other hand, it is known that the loop with two $t$-channel fermions
may contain a double logarithm (see~\cite{GGLF,KLip}). One logarithm comes
from the transverse ($Q_{T}$) integration, while another logarithm (in the
real part of the amplitude) originates from the $dm^{2}/m^{2}$ integral over
the (virtual) mass of the upper $s$-channel particle in the loop (assuming
that the contour of the Feynman integral is closed on the pole corresponding
to the lower $s$-channel particle).  In our kinematics, where a $Z$ boson is
emitted in the centre of the rapidity gap interval, we obtain a logarithm
when the mass, $m$, runs from
$m^{2}=\textrm{max}\{Q_{T}^{2},\sqrt{\hat{s}M_{ZT}^{2}}\}$ up to
$m^{2}=\hat{s}$ (here $\hat{s}$ is the incoming parton energy squared and
$M_{ZT}^{2}=M_{Z}^{2}+|p_{TZ}^{2}|$).  That is, the mass integral gives
$\int\frac{dm^{2}}{m^{2}}\lesssim\frac{\Delta\eta}{2}$.  The $Q_{T}$
integration does not give a logarithm in the case of Fig.~\ref{screenq}a, but
for the amplitudes corresponding to Fig.~\ref{screenq}b,c, a logarithmic
integral appears in the domain $p_{T\textrm{jet}}^{2}\ll Q_{T}^{2}\ll
\hat{s}^{2}/4$. Thus from the Fig.~\ref{screenq}b,c loop integrals we may
expect a $\frac{3}{8}(\Delta\eta)^{2}$ enhancement. However, with our large
rapidity gap ($\Delta\eta\simeq 6$) the whole factor --
$[\frac{\alpha_{S}}{\pi}\frac{3}{8}(\Delta\eta)^{2}e^{-\Delta\eta/2}]^{2}
=0.45(\frac{\alpha_{S}}{\pi})^{2}\sim 10^{-3}$ is very small. Besides this,
after the parton-level cuts described in Section~\ref{subsec:dist} are
applied, the original parton-level contribution of the diagrams with a
($t$-channel) quark exchange is strongly suppressed.  Therefore we neglect
these contributions.

\subsection{Soft Survival Probability}
\label{subsec:ssp}
Returning to our original processes, we keep now only the graphs with either
vector boson or two gluon (Fig.~\ref{screen}) exchange across the rapidity
gaps, and multiply the corresponding parton-level cross sections by the
appropriate gap survival probability $\hat{S}^{2}$. However, as we now
require there to be no hadrons (even with a rather low $p_{T}$) in the gap
interval, we have to take account of any soft interactions of the spectator
partons.\par Instead of using a Monte Carlo simulation, it is better to
choose a model based on Regge (Pomeron) theory tuned to describe soft
interaction data at high energies. We will use the model of Ref.~\cite{18}.
This is based on the two-channel eikonal formalism, which reproduces all the
main features of the soft ($\sigma_{tot}$, $d\sigma_{el}/dt$) cross section
data in the ISR--Tevatron energy range. Recall that the two channels of the
eikonal correspond to two eigenstates which have different absorptive cross
sections (i.e. different rescattering probabilities).  Assuming the same
(momentum and spatial) distributions of quarks and gluons in both components
of the incoming proton wavefunction (that is, in both eigenstates of the
eikonal) the model predicts for all our processes $\hat{S}^{2}=0.1$ at LHC
energies. In other words, by requiring gaps at the hadron level we decrease
the cross section by an order of magnitude. At first sight, the gap survival
probability $\hat{S}^{2}=0.1$ reflects the rescattering of soft spectator
partons and should, therefore, be universal for any process which has a
gap\footnote{The only difference may be caused by the Sudakov-like form
  factor that accounts for the absence of QCD gluon bremsstrahlung in a
  specific hard subprocess.}. However, this is not completely true. First,
the value of $\hat{S}^{2}$ depends on the spatial distribution of parton
spectators and therefore on the characteristic impact parameter ($b_{T}$)
difference between the two colliding protons~\cite{18,19}. For example, in
the case of exclusive Higgs boson production, $pp\rightarrow p+H+p$ via
photon-photon fusion, the transverse momenta of the photons are very small.
Hence the impact parameter $b_{T}$ is very large. The probability of soft
rescattering in such a highly peripheral collision is small, and the value of
$\hat{S}^{2}$ ($\sim 0.9$) is close to one~\cite{9,11}. Secondly, there is a
difference in the momentum distributions of partons in a different
(diffractive eigenstate) component of the incoming proton wavefunction; it is
reasonable to expect that the component with a smaller cross section contains
more valence quarks (and `hard' large-$x$ partons), whereas the component
with a larger cross section has more low-$x$ gluons. This possibility was
discussed in~\cite{19}. In such an approach, the model describes the
breakdown of factorisation, in that there is about a factor $10$ difference
between the `effective' Pomeron structure functions measured in diffractive
deeply inelastic interactions at HERA and diffractive high-$E_{T}$ dijet
hadroproduction at the Tevatron~\cite{23}\footnote{The difference is
  explained simply by the fact that the gap survival factor is
  $\hat{S}^{2}\sim 0.1$ for proton-antiproton collisions, whereas
  $\hat{S}^{2}\simeq 1$ in deep inelastic scattering.}.\par In the present
context, as the background $b\bar{b}$-pairs are produced predominantly in
gluon-gluon collisions, the gap survival probability for the QCD background
is a little smaller than for $Z$($H$)-boson production via vector boson
fusion where we deal with incoming quark-quark interactions (see
\Figs{h}{zew}). Using the formalism of Ref.~\cite{19} we obtain for the
kinematics (cuts) described in Section~\ref{sec:partonchar},
\begin{equation}
\label{surfacts}
\hat{S}_{Z}^{2}=0.31;\qquad\hat{S}_{H}^{2}=0.31;\qquad\hat{S}_{QCD b\bar{b}}^{2}=0.27. 
\end{equation}
These survival factors are much larger than in the original model~\cite{18}
because for the case considered here, of large rapidity gaps and large jet
transverse momenta, we select mainly fast incoming partons and valence quarks
which belong to the second component of the proton wavefunction. This
component has a smaller absorptive cross section\footnote{Under the extreme
  hypothesis that all valence quarks belong to the second (low
  $\sigma_{abs}$) component while gluons and sea quarks are concentrated in
  the first component (with a larger cross section) we get
\begin{equation}
\label{KKMR1}
\nonumber \hat{S}_{H}^{2}=\hat{S}_{Z}^{2}=0.26
\qquad\textrm{and}\qquad\hat{S}_{QCD b\bar{b}}^{2}=0.10. 
\end{equation}}
In this case the QCD background is additionally suppressed $2.5$ times.  Note
that both versions of the model~\cite{19} describe the diffractive dijet CDF
data~\cite{23} well enough. On the other hand, in processes with large
rapidity gaps at the LHC the uncertainty in the soft survival factor
$\hat{S}^{2}$ may be rather large. It will therefore be important to study
such a process experimentally. A promising way to study the survival
probability $\hat{S}^{2}$ in different components of the incoming proton wave
function (i.e. the dependence of $\hat{S}^{2}$ on the $p_{T\textrm{jet}}$ and
rapidity cuts) is to measure QCD dijet production with rapidity gaps on
either side of the dijet pair. Here the cross section is much larger
(especially for gluon-gluon induced dijets) and it is easy to study the gap
survival factor $\hat{S}^{2}$ under the various kinematic conditions: $p_{T}$
of the fast (large $\eta$) jets, size of the rapidity gaps, dijet mass, etc.
In this way we can emphasise the r\^{o}le of the incoming valence quarks, sea
quarks or gluons in different $x$ and scale $\mu^{2}\sim p_{T}^{2}$ domains,
and hence choose the configuration where one or other component of the
wavefunction dominates.\par Note that, depending on the jet-finding
algorithm, some soft hadrons may or may not be attributed to a particular
$b$-jet. Therefore, one has to be more specific in the definition of the
rapidity gap on the hadronic level in the presence of the high-$p_{T}$ jets.
It looks plausible to select the gap by the requirement not to have hadrons
within the gap range, apart from the cones of a fixed size $\Delta R\sim 1$
around the jet directions. In a real life experiment, jet-finding algorithms
should be utilised in optimising the value of $\Delta R$. Soft survival
factors $\hat{S}^{2}$ are practically independent of the $\Delta R$ value at
$\Delta R\le 2$.

\section{Results}
\label{sec:results}
\begin{figure}[!ht]
\hspace{15mm}
\scalebox{0.45}{\includegraphics[angle=-90]
{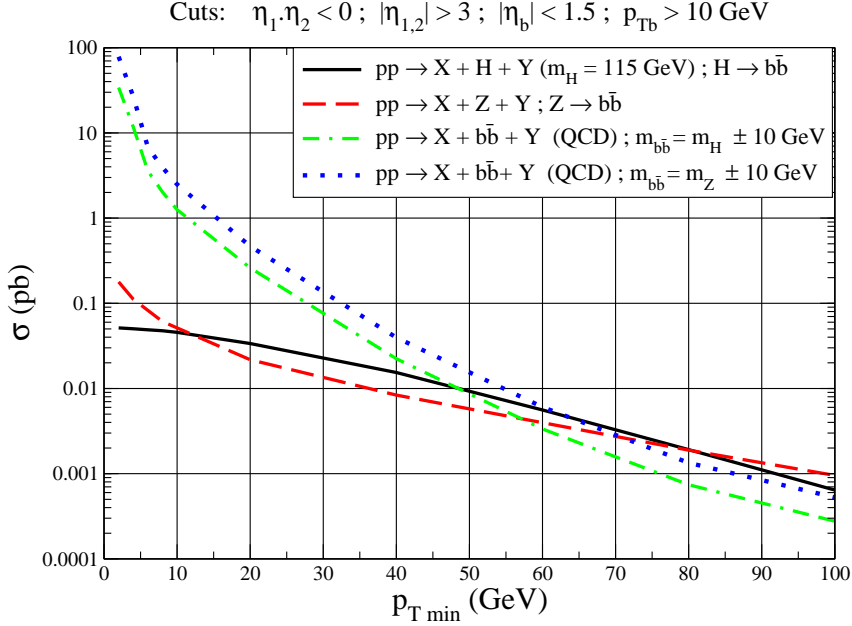}}
\caption{Hadron-level cross sections at $\sqrt{s} = 14$ TeV for inclusive Higgs and $Z$ production with subsequent decay to $b\bar{b}$ and their respective QCD $b\bar{b}$ backgrounds. Cuts are applied at the parton level as discussed in the text.}
\label{final_plot}
\end{figure}
\Fig{final_plot} shows the cross sections after hadronisation for central
production of Higgs or $Z$ with rapidity gaps and subsequent decay to
$b\bar{b}$ as a function of $\ptmin$ of the proton remnant jets. It also
shows the expected background of QCD $b\bar{b}$ events that display the same
kinematical configuration. These are calculated using as a starting point the
parton level cross sections after application of cuts, namely
Figure~\ref{cut_sig_H} for Higgs production and Figure~\ref{cut_sig_Z} for
$Z$ production. The QCD-induced cross sections (both the QCD $Z$ production
of \Fig{zqcd}d and the direct QCD $b\bar{b}$ production of \Fig{hbk}) are
then multiplied by the probability to screen out the colour octet
contribution for the relevant initial state of either $qq$, $qg$ or $gg$
(\Eqn{prob}). To take into account the fact that the $b\bar{b}$ pair in the
background processes can only be produced in the colour singlet state the
ordinary $gg\rightarrow b\bar{b}$ cross section is replaced by the pure
singlet cross section, \Eqn{singlet}. Finally both the signals and
backgrounds are multiplied by the relevant soft survival probability of
\Eqn{surfacts}.\par We see that, as long as we stay away from the low
$\ptmin$ region, the signal for Higgs production is comparable with the QCD
background, even exceeding it above $\ptmin=50$~GeV. The cross section for
$Z$ production with rapidity gaps is less than that for Higgs production over
most of the plot. This is because the branching fraction to $b\bar{b}$ is
much lower than for the Higgs. Exceptions to this occur at low $\ptmin$ where
the effect of the infrared singularity makes its presence felt and at very
high $\ptmin$, explained by the fact that the parton-level Higgs cross
section falls more rapidly, as shown in \Fig{totx}. The backgrounds show an
extremely strong dependence on $\ptmin$, falling by five orders of magnitude
as one varies $\ptmin$ from $2$~GeV to $100$~GeV. The QCD $b\bar{b}$
background with the invariant mass of the $b\bar{b}$ pair taken around the
$Z$ mass is approximately $80\%$ higher than that evaluated around the Higgs
mass.\par It should be noted that in these calculations we have taken
$\alpha_S\equiv\alpha_S(M_{Z}^2)$. An argument could also be made that the
characteristic scale should be that of the transverse momenta of the forward
jets, i.e. $\alpha_S\equiv\alpha_S(\ptmin^2)$. This would affect the
$\mathcal{O}(\alpha_S^4)$ backgrounds in such a way as to increase them by
approximately $30\%$ if we take the typical $p_{T}$ to be $40$~GeV.\par
Let us emphasise that up to now we have not addressed the experimental
issues. In particular, the predictions given in Fig.~\ref{final_plot} should
be modified to account for the $b$-tagging efficiency $\varepsilon_{b}$. This,
in turn, is correlated with the probability ${\cal P}(g,q/b)$ to misidentify
a gluon (or a light quark) jet as a $b$-jet. Recall that the rate of the
$gg$-dijets exceeds the $b\bar{b}$-yield by two orders of magnitude. As
discussed in \cite{dkmoz}, it is feasible to expect for the two $b$-jets
$(\varepsilon_{b})^2=0.6$ with ${\cal P}(g,q/b)=0.01$. 
\section{Conclusions}
\label{sec:conc}
The weak boson fusion mechanism can provide a promising way to detect a light
Higgs boson at the LHC, see for example~\cite{17,zknr,28}. The selection of
events with large rapidity gaps and energetic large $p_T$ (quark) jets in the
forward and backward directions allows the suppression of the QCD $b\bar{b}$
background down to a level comparable to the signal. Therefore, it becomes
feasible to observe a light Higgs boson via its main $H\rightarrow b\bar{b}$
decay mode in addition to the usually discussed $\tau\tau$ and $WW^*$
channels, see for example~\cite{dkmoz}.\par The cross section for the
production of a $115$ GeV Higgs boson in association with rapidity gaps at
the LHC is expected to be about $15$ fb (for $p_T > 40$~GeV). Therefore, for
an integrated luminosity of $30 \textrm{ fb}^{-1}$ planned for the first two
or three years of LHC running, about 400 events can be observed.\par Note
that our cuts were not finally optimised for the particular ATLAS/CMS
conditions. Thus, the significance of the signal may be improved by allowing
asymmetric configurations with some minimal $\Delta\eta$ between the
high-$p_{T}$ jets instead of the requirement $|\eta_{1,2}|\ge 3$,
$\eta_{1}\cdot\eta_{2}\le 0$. Such a kinematical choice was considered, for
instance, in Ref.~\cite{17}. It is shown that this will noticeably improve
the significance of the signal.\par An important ingredient in the evaluation
of both the signal and the background in the $b\bar{b} + 2$ forward jet
events is the soft survival factor $\hat{S}^2$, defining the probability that
the gaps survive the soft $pp$-scattering. Recall that though this factor can
be computed within the framework of existing models for soft rescattering, it
is always unwise to rely on the precision of models based on soft physics.
Fortunately, the soft survival factor for the gaps surrounding $WW\rightarrow
H$ fusion can be monitored experimentally by observing the closely related
central production of a $Z$ boson with the same rapidity gap and jet
configuration~\cite{26,27}.\par As was emphasised in~\cite{27,28,RZ}, the
$\tau\tau$ and $WW^*$ decay channels with rapidity gap kinematics can give a
rather high significance for the observation of a light Higgs. In the
$\tau\tau$ case the main background results from the tail in the $\tau\tau$
mass distribution generated by the $Z\rightarrow\tau^{+}\tau^{-}$ decay.
Again, the experimental observation of $Z$ boson central production allows
one to control and monitor such a background.\par It is worthwhile to mention
that the experimental determination of the gap survival factor in the
processes under consideration is interesting in its own right, since it
provides important information on the incoming proton wavefunction. Note that
since the incoming partons in the subprocess $qq\rightarrow q + (b\bar{b}) +
q$ are rather hard, the factor $\hat{S}^2$ depends on the model assumptions
more sensitively than, for example, in the exclusive diffractive production
case $pp\rightarrow p + b\bar{b} + p$, see~\cite{11,18}. As was demonstrated
in Section~\ref{subsec:ssp} (see footnote$^9$) the results strongly depend on
how the partons in the proton are distributed between the different
diffractive eigenstates. Currently our information on these distributions is
rather limited.\par This paper concentrates on the detailed analysis of
central $Z$ boson production accompanied by rapidity gaps on either side and
two forward jets at the LHC. The QCD background processes for $Z + 2$ jet
production in the rapidity gap environment are addressed in detail. We
evaluate the soft survival factors $\hat{S}^2$ for various processes under
consideration. Finally, we note that it will be important to extend our work
by incorporating a realistic Monte Carlo simulation, which will allow
detector simulation to be included. We believe that the results presented in
this paper make such an effort worthwhile.

%%%%%%%%%%%%%%%%%%%%%%%%%%%%%%%%%%%%%%%%%%%%%%%%%%

\section*{Acknowledgments}
We thank A.~De Roeck, A.~D.~Martin and R.~Orava for useful discussions. One
of us (VAK) thanks the Leverhulme Trust for a Fellowship. This work was
partially supported by the UK Particle Physics and Astronomy Research
Council, by the Russian Fund for Fundamental Research (grants 01-02-17095 and
00-15-96610) and by the EU Framework TMR programme, contract FMRX-CT98-0194
(DG12-MIHT).

%%%%%%%%%%%%%%%%%%%%%%%%%%%%%%%%%%%%%%%%%%%%%%%%%%%%%%%%%%%

% A useful Journal macro
\def\Journal#1#2#3#4{{#1} {\bf #2}, #3 (#4)}

% Some useful journal names
\def\NCA{Nuovo Cimento}
\def\NIM{Nucl. Instrum. Methods}
\def\NIMA{Nucl. Instrum. Methods {\bf A}}
\def\NPB{Nucl. Phys. {\bf B}}
\def\PLB{Phys. Lett. {\bf B}}
\def\PRL{Phys. Rev. Lett. }
\def\PRD{Phys. Rev. {\bf D}}
\def\ZPC{Z. Phys. {\bf C}}

\end{document}